\documentclass[aps,prl,twocolumn,superscriptaddress,groupedaddress,showkeys]{revtex4}  
\usepackage{graphicx}  
\usepackage{dcolumn}   
\usepackage{bm}        
\usepackage{amssymb}   
\usepackage{amsmath}
\usepackage{xcolor}
\usepackage{amsthm}
\newtheorem{theorem}{Theorem}
\newcommand{\be}{\begin{equation}}\newcommand{\ee}{\end{equation}}
\newcommand{\bea}{\begin{eqnarray}}\newcommand{\eea}{\end{eqnarray}}
\newcommand{\brr}{\begin{array}}\newcommand{\err}{\end{array}}
\newcommand{\bit}{\begin{itemize}}\newcommand{\eit}{\end{itemize}}
\newcommand{\ben}{\begin{enumerate}}\newcommand{\een}{\end{enumerate}}

\newcommand{\bbm}{\begin{bmatrix}}\newcommand{\ebm}{\end{bmatrix}}
\newcommand{\ba}{\begin{array}}
\newcommand{\ea}{\end{array}}
\newcommand{\G}{\textbf}

\newtheorem{mydef}{Definition}
\newtheorem{Lemma}{Lemma}
\newcommand{\bd}{\begin{mydef}} \newcommand{\ed}{\end{mydef}}
\newcommand{\bthe}{\begin{theorem}} \newcommand{\ethe}{\end{theorem}}
\newcommand{\ble}{\begin{Lemma}} \newcommand{\ele}{\end{Lemma}}

\newcommand{\dr}{\mathrm{d}}

\definecolor{darkred}{rgb}{.8,0,0}

\definecolor{darkblue}{rgb}{0,0,.7}

\def\ha{\frac{1}{2}}

\def\intx{\int \!\!\mathrm{d}^3 x}
\def\intk{\int \!\!\mathrm{d}^3 k}

\def\lab{\label}\def\lan{\langle}
\def\lf{\left}

\def\non{\nonumber}\def\pa{\partial}\def\ran{\rangle}

\def\ri{\right}
\def\al{\alpha}\def\bt{\beta}\def\ga{\gamma}
\def\de{\delta}\def\De{\Delta}

\def\si{\sigma}
\def\om{\omega}

\def\1{{_{1}}}\def\2{{_{2}}}

\def\noHe0{:\;\!\!\;\!\!:H_e(0):\;\!\!\;\!\!:}
\def\noHm0{:\;\!\!\;\!\!:H_\mu(0):\;\!\!\;\!\!:}

\def\lab{\label}
\def\lan{\langle}
\def\lf{\left}

\def\non{\nonumber}
\def\pa{\partial}\def\ran{\rangle}

\def\ri{\right}

\def\al{\alpha}\def\bt{\beta}\def\ga{\gamma}
\def\de{\delta}\def\De{\Delta}

\def\si{\sigma}
\def\om{\omega}

\def\1{{_{1}}}\def\2{{_{2}}}

\begin{document}

\title{Flavor-Energy uncertainty relations for neutrino oscillations in quantum field theory}
\author{Massimo~Blasone}
\email{blasone@sa.infn.it}
\affiliation{Dipartimento di Fisica, Universit\`a di Salerno, Via Giovanni Paolo II, 132 84084 Fisciano, Italy \& INFN Sezione di Napoli, Gruppo collegato di Salerno, Italy}
\author{Petr Jizba}
\email{p.jizba@fjfi.cvut.cz}
\affiliation{FNSPE, Czech Technical
University in Prague, B\v{r}ehov\'{a} 7, 115 19 Praha 1, Czech Republic\\}
\affiliation{ITP, Freie Universit\"{a}t Berlin, Arnimallee 14,
D-14195 Berlin, Germany}
\author{Luca~Smaldone}
\email{lsmaldone@sa.infn.it}
\affiliation{Dipartimento di Fisica, Universit\`a di Salerno, Via Giovanni Paolo II, 132 84084 Fisciano, Italy \& INFN Sezione di Napoli, Gruppo collegato di Salerno, Italy}
%
\vspace{3mm}

\begin{abstract}
In the context of quantum field theory, we derive  flavor-energy uncertainty relations for neutrino oscillations.
By identifying the non-conserved flavor charges with the ``clock observables'',  we arrive at the Mandelstam--Tamm version
of time-energy uncertainty relations. In the ultra-relativistic limit
these  relations  yield the well known condition for neutrino oscillations.
Ensuing non-relativistic corrections to the latter are explicitly evaluated.
The analogy among flavor states and unstable particles and a novel interpretation of our uncertainty relations, based on the unitary inequivalence of Fock spaces for flavor and massive neutrinos, are also discussed.
\end{abstract}


\maketitle

\noindent{\em {Introduction}.}~---~
Neutrino mixing and oscillations represent one of the most pressing challenges of modern theoretical and experimental particle physics.
They were first introduced  by Pontecorvo~\cite{Pontecorvo} in a close analogy with the phenomenon of Kaon oscillations~\cite{GelPai},
and subsequently confirmed in a number of experimental settings~\cite{neutrinoexp}.
While the quantum mechanical (QM) description~\cite{Kayser,BilPet,giuntibook}
is quite successful in tackling high-energy features of neutrino oscillations,
the corresponding quantum field theoretical (QFT) description (which could tackle also the low-energy behavior)
is still controversial~\cite{giuntibook,NeutPheno,qftmixing,qftmixing2}.

In particular, a non-perturbative study of flavor states shows that field mixing is not the same as the wave-function (i.e., first-quantized) mixing~\cite{qftmixing}. In fact, the \emph{flavor vacuum} is structurally a condensate similar to the BCS vacuum~\cite{qftmixing,Chang}.
From this, the corrections to the standard neutrino oscillation formula can be derived~\cite{BHV99}.
It should be stressed that in this analysis a central r\^{o}le is played by non-conserved flavor-charges implied by the Noether's theorem~\cite{ChargesBJV}. Flavor states are defined as eigenstates of such charges and have the form of  $SU(2)$ generalized coherent states~\cite{Perelomov}.
In fact, according to Standard Model (SM), the flavor charge is always conserved -- at tree level -- in the production and detection processes, a feature which is violated by the usual Pontecorvo flavor states~\cite{WeakDecay}.
In addition, the exact flavor states cannot be generally phrased as a simple superposition of mass eigenstates, because of the unitary inequivalence of mass and flavor Hilbert spaces~\cite{qftmixing}.

In~Ref.~\cite{Bil}, it was shown that for neutrino oscillations described in terms of Pontecorvo states, the Mandelstam--Tamm time-energy uncertainty relations (TEUR)~\cite{ManTam} reduce to  the known
condition for neutrino oscillations~\cite{Kayser}. However, this result was obtained in the context of
standard  perturbative treatment of neutrino flavor states.

In this Letter we employ the full  QFT framework to derive the {\em flavor--energy} uncertainty relations (FEUR).
By identifying the non-conserved flavor charge with the ``clock observable''  we find from the latter the Mandelstam--Tamm version
of TEUR. Our approach is valid at all energy scales and the conventional results of Bilenky {\em et al.}~\cite{Bil}
are recovered in the ultra-relativistic limit.
Moreover, by exploiting the analogy between flavor neutrinos and unstable particles~\cite{Bhat}
we find that for, an exact neutrino flavor state, an inherent energy uncertainty
arises from TEUR. Although our discussion is, for simplicity's sake, confined to
two flavors only, the results obtained can be easily extended to three flavors including $CP$-violation.

{\em {Neutrino mixing and oscillations in QFT}.}~---~
Let us consider a weak decay  $W^+\rightarrow e^+ + \nu_e$. The relevant part of SM Lagrangian is
$\label{Lagrangian} \mathcal{L}=\mathcal{L}_0+\mathcal{L}_{int}$ with
\bea
&&\mbox{\hspace{-2mm}}{\cal L}_{0}  =  \overline{\nu}  \lf( i \ga_\mu \pa^\mu - M_{\nu} \ri)\nu \, + \, \overline{l} \lf( i \ga_\mu \pa^\mu - M_{l} \ri) l  \nonumber \, , \\[2mm]
 &&\mbox{\hspace{-2mm}}{\cal L}_{int}  =  \frac{g}{2\sqrt{2}}
\lf [ W_{\mu}^{+}\,
\overline{\nu}\,\gamma^{\mu}\,(1-\gamma^{5})\,l +
h.c. \ri] \, ,
\label{L-interact}
\eea
where $\nu = \lf(\nu_e, \nu_\mu \ri)^{{T}}  , \, l = \lf(e, \mu \ri)^T$, and
\bea
&&\mbox{\hspace{-5mm}} M_{\nu}\,=\,  \lf(\ba{cc}m_e & m_{e\mu}
\\ m_{e\mu} & m_\mu\ea\ri), \;\;\;\;\;
M_l\,=\,  \lf(\ba{cc}\tilde{m}_e &0 \\
0 & \tilde{m}_\mu\ea\ri) .
\eea
The Lagrangian $\mathcal{L}$ is invariant under the global $U(1)$ transformations
$\nu \rightarrow e^{i \alpha} \nu$ and $l \rightarrow e^{i \alpha} l$
leading to the conservation of the total  flavor charge $Q_{l}^{tot}$ corresponding to the lepton-number conservation~\cite{BilPet}. This can be written in terms of the flavor charges for neutrinos and charged leptons~\cite{ChargesBJV}
\be
Q_{l}^{tot} =  \sum_{\si=e,\mu} Q_\si^{tot}(t) \,,\quad   Q_{\si}^{tot} (t) = Q_{\nu_{\si}}(t) + Q_{\si}(t)\,,
\ee
with
\bea
&&\mbox{\hspace{-8.5mm}}Q_{e} =  \intx \,
e^{\dag}(x)e(x) \,, \;\; Q_{\nu_{e}} (t) =  \intx \,
\nu_{e}^{\dag}(x)\nu_{e}(x)\,,
\nonumber \\ [1mm]
&&\mbox{\hspace{-8.5mm}}Q_{\mu} =   \intx \,
 \mu^{\dag}(x) \mu(x)\,, \;\; Q_{\nu_{\mu}} (t)= \intx \, \nu_{\mu}^{\dag}(x) \nu_{\mu}(x)\,  .
 \label{QflavLept}
\eea
The above charges can be derived via Noether's theorem~\cite{ChargesBJV}  from the Lagrangian~\eqref{L-interact}. Note the time dependence of the neutrino charges, due to the non-diagonal mass matrix $M_{\nu}$.

 By observing that  $[\mathcal{L}_{int}({\bf x},t),Q_\si^{tot}(t)]=0$, we see that a neutrino flavor state is well defined in the production vertex
as an eigenstate of the corresponding  flavor charge~\cite{Chargeei}.

The mixing relations for neutrino fields are
\bea
\left(
  \begin{array}{c}
    \nu_e (x) \\
    \nu_{\mu}(x) \\
  \end{array}
\right) \ = \ \left(
                \begin{array}{cc}
                  \cos\theta & \sin\theta \\
                  -\sin\theta & \cos\theta \\
                \end{array}
              \right)\left(
                       \begin{array}{c}
                        \nu_1 (x)\\
                         \nu_2 (x) \\
                       \end{array}
                     \right), \label{PontecorvoMix1}
\eea
with $\tan 2 \theta = 2 m_{e\mu}/(m_\mu-m_e)$. The fields with definite masses have the usual mode expansion
\begin{eqnarray}
\mbox{\hspace{-15mm}}&&\nu _{j}(x) =  \sum_r \,  \int \!\! \frac{\dr^3 k}{(2 \pi)^{\frac{3}{2}}}\!\! \ \,
\left[ u_{{\bf k},j}^{r}(t) \, \alpha _{{\bf
k},j}^{r}(t) \right. \nonumber \\[1mm]
&&\mbox{\hspace{13mm}}\left.+  \ v_{-{\bf k},j}^{r}(t)  \beta _{-{\bf k},j}^{r\dagger
}(t)\right]  e^{i{\bf k}\cdot {\bf x}} \, ,  \quad j=1,2 \, .
\label{fieldex}
\end{eqnarray}
where
\begin{eqnarray}
\left(
  \begin{array}{c}
    m_e \\
    m_\mu \\
  \end{array}
\right) \ = \ \left(
                \begin{array}{cc}
                  \cos^2 \theta & \sin^2 \theta \\
                  \sin^2 \theta & \cos^2 \theta \\
                \end{array}
              \right)\left(
                       \begin{array}{c}
                         m_1 \\
                         m_2 \\
                       \end{array}
                     \right).
\end{eqnarray}
Eq. (\ref{PontecorvoMix1}) can be equivalently rewritten as~\cite{qftmixing}
\bea
\label{MixingRel1}
\nu_\si(x)=
G_{\theta}^{-1}(t)\, \nu_j (x) \,  G_{\theta}(t) \, , \label{MixingRel2}
\eea
with $(\si,j)=(e,1),(\mu,2)$ and  $G_\theta(t)$  given by
\be
\mbox{\hspace{-2mm}}G_{\theta}(t)=\exp\left[\theta\intx \, \left(\nu_1^{\dagger}(x)\nu_2(x)-
\nu_2^{\dagger}(x)\nu_1(x)\right)\right].
\label{MixGen}
\ee
From (\ref{fieldex}) and (\ref{MixingRel1}) it follows that flavor fields are:
\begin{eqnarray}
\mbox{\hspace{-20mm}}&&\nu _{\si}(x) \ = \ \sum_r \,  \int \!\! \frac{\dr^3 k}{(2 \pi)^{\frac{3}{2}}}\!\! \ \,
\left[ u_{{\bf k},j}^{r}(t) \, \alpha _{{\bf
k},\si}^{r}(t)\right.\nonumber \\[2mm]
&&\mbox{\hspace{10mm}}+ \left. v_{-{\bf k},j}^{r}(t) \, \beta _{-{\bf k},\si}^{r\dagger
}(t)\right] \, e^{i{\bf k}\cdot {\bf x}} \, ,
\label{fieldex1}
\end{eqnarray}
with $(\si,j) \ = \ (e,1), (\mu,2)$, and flavor ladder operators  given by~\footnote{Here we choose a Lorentz frame so that $\textbf{k}=(0,0,|\textbf{k}|)$.}
\bea\label{flava}
\begin{pmatrix}
\alpha^{r}_{{\bf
k},\si}(t)\\
\beta^{r}_{-{\bf k},\si}(t)
\end{pmatrix}
\ = \
G_{\theta}^{-1}(t)\,
\begin{pmatrix}
\alpha^{r}_{{\bf
k},j}(t)\\
\beta^{r}_{-{\bf k},j}(t)
\end{pmatrix}
\,  G_{\theta}(t) \, .
\eea
The vacuum for massive fields, $|0 \rangle_{1,2}$, is defined as:
\be
\alpha^{r}_{{\bf k},j} |0 \rangle_{1,2} \ = \ \bt^{r}_{-{\bf k},j} |0 \rangle_{1,2} \ = \ 0 \, \qquad j=1,2 \, ,
\ee
and is not left-invariant under the action of $G_\theta(t)$:
\be
\label{flavvac} |0(t) \rangle_{e,\mu} = G^{-1}_{\theta}(t)\;
|0 \rangle_{1,2}\;.
\ee
$|0(t) \rangle_{e,\mu}$ is called \emph{flavor vacuum}, and one can easily verify that it is annihilated by the flavor operators
defined in Eq.~\eqref{flava}. Moreover, one can prove~\cite{qftmixing} that
\be \label{ineqrep}
\lim_{V \rightarrow \infty}{}_{1,2} \lan 0|0(t) \rangle_{e,\mu} =  \lim_{V \rightarrow \infty} e^{\frac{V}{(2\pi)^3}\intk (1-\sin^2 \theta|V_\G k|^2)^2}  =  0 \, ,
\ee
where
\bea
&&\mbox{\hspace{-9mm}}|V_\G k|  = \sqrt{\frac{|\G k|}{4 \om_{\G k,1}\om_{\G k,1}}}
\lf(\sqrt{\frac{\om_{\G k,2}+m_2}{\om_{\G k,1}+m_1}}-\sqrt{\frac{\om_{\G k,1}+m_1}{\om_{\G k,2}+m_2}}\ri)\!\!,
\eea
i.e. flavor and massive fields belong to unitarily inequivalent representations of the anticommutation relations~\cite{qftmixing}.\\

{\em {Mandelstam--Tamm TEUR}.}~---~
%
Mandelstam--Tamm version of TEUR is formulated as~\cite{ManTam}
%
%
\be \label{teunc}
\Delta E \, \Delta t \, \geq \frac{1}{2} \, ,
\ee
where
\be
\Delta E \equiv \si_H \, \qquad \Delta t \equiv \si_O/\lf|\frac{\dr \lan O(t) \ran}{\dr t}\ri| \, .
\label{teunc1}
\ee
Here $O(t)$ represents the ``clock observable'' whose dynamics quantifies temporal changes in a system
and $\Delta t$ is the characteristic time interval over which the mean value of $O$ changes by a standard deviation.

TEUR (\ref{teunc})-(\ref{teunc1}) is typically applied to the study of unstable particles, see, e.g.~\cite{Schumacher}. Calling $|\phi(t)\ran$ an eigenstate
of the projection operator $P_\phi(t) \ = \ |\phi(t)\ran \lan \phi(t)|$
%
describing an unstable particle state, one gets~\cite{Bhat}
\be \label{decineq}
\lf|\frac{\dr \mathcal{P}_\phi(t)}{\dr t}\ri|  \,\leq \, 2 \Delta E \,\sqrt{\mathcal{P}_\phi(t)(1-\mathcal{P}_\phi(t))} \, .
\ee
Here $\mathcal{P}_\phi(t)$ is the survival probability
\be
\mathcal{P}_\phi(t) \ = \ |\lan \phi(t)|\phi\ran|^2 \, ,
\ee
and $|\phi\ran$ is the (Heisenberg representation) state of the system prepared at
$t=0$.

The r.h.s. of~(\ref{decineq}) has a maximum when $\mathcal{P}_\phi(t)=\ha$, which is satisfied with $t=T_h$. Thus, we have
\be \label{simpdis}
\Delta E  \ \geq \ \lf|\frac{\dr \mathcal{P}_\phi(t)}{\dr t}\ri| \, .
\ee
Because for decaying particles $\mathcal{P}_\phi(0)=1$, $\mathcal{P}_\phi(\infty)=0$ and $\mathcal{P}_\phi(t)$ is a monotonically decreasing,
we can integrate the inequality~(\ref{decineq}), obtaining
\be \label{integratedte}
\De E \, T \geq \ha \lf[\frac{\pi}{2}-\arcsin\lf(2 \mathcal{P}_\phi(T)-1 \ri)\ri] \, .
\ee
From this, one can derive an explicit form of TEUR for unstable particles~\cite{Bhat}
\be \label{bhatin}
\De E \,T_h \ \geq \ \frac{\pi}{4} \, .
\ee

One can use a similar line of reasonings to arrive at the TEUR for the neutrino oscillations~\cite{Bil}.
However, in order to remain as close as possible to the full QFT treatment, we employ a different strategy.
We start by considering the number operator for flavor neutrinos:
\be {
\tilde{N}_\si(t) \ = \ \sum_{\G k,r} \tilde{\al}^{r\dag}_{\G k, \si}(t) \tilde{\al}^r_{\G k, \si}(t) \,, \qquad \si=e,\mu \, ,}
\ee
where
\begin{eqnarray}
\left(
  \begin{array}{c}
    {\tilde{\al}^{r}_{\G k,e}} \\
    {\tilde{\al}^{r}_{\G k,\mu}} \\
  \end{array}
\right) \ \equiv \
\left(
  \begin{array}{cc}
    \cos \theta & \sin \theta \\
    -\sin \theta & \cos \theta \\
  \end{array}
\right)
\left(
  \begin{array}{c}
    \al^r_{\G k,1} \\
    \al^r_{\G k,2} \\
  \end{array}
\right).
\end{eqnarray}
These relations are just approximations of
the exact ones \eqref{flava} in the ultra-relativistic limit.
Defining the \emph{Pontecorvo flavor state} as
\be\lab{pontstate}
|\nu^r_{\G k,\si}\ran_P \ \equiv \ \tilde{\al}^{r\dag}_{\G k,\si}|0\ran_{1,2} \,,
\ee
{one gets the usual relations among flavor and massive neutrino states}
\begin{eqnarray}
\left(
  \begin{array}{c}
    |\nu^r_{\G k,e}\ran_P \\
    |\nu^r_{\G k,\mu}\ran_P \\
  \end{array}
\right) \ = \ \left(
                \begin{array}{cc}
                  \cos \theta & \sin \theta \\
                  -\sin \theta & \cos \theta \\
                \end{array}
              \right) \left(
                        \begin{array}{c}
                          |\nu^r_{\G k,1}\ran \\
                          |\nu^r_{\G k,2}\ran \\
                        \end{array}
                      \right).
\label{ponte1}
\end{eqnarray}
The standard oscillation formula~\cite{Pontecorvo}  can be found by taking the expectation value of the number operator 
over the corresponding Pontecorvo flavor state
\begin{eqnarray}\label{pontosc}
\mbox{\hspace{-3mm}}\mathcal{P}_{\!\si\rightarrow \si}(t)\!=\! \lan \tilde{N}_\si(t) \ran_\sigma  \!=\!  1\! -\! \sin^2(2\theta)\sin^2 \!\lf(\!\frac{\om_{\G k, 1}-\om_{\G k,2}}{2}t\!\ri)\!,
\end{eqnarray}
where $\langle \cdots\rangle_\sigma = {}_P\lan \nu^r_{\G k,\si}| \cdots |\nu^r_{\G k,\si}\ran_P$. By setting $O(t)={\tilde{N}_\si(t)}$
in (\ref{teunc1}) and taking into account that
\bea \non
\si^2_N & = &  \lan \tilde{N}^2_\si(t) \ran_\sigma \ - \ \lan \tilde{N}_\si(t) \ran_\si^2 \\[1mm]
& = & \ \mathcal{P}_{\si\rightarrow \si}(t)\lf(1-\mathcal{P}_{\si\rightarrow \si}(t)\ri) ,
\eea
one gets
\be \label{neutunqm}
\lf|\frac{\dr \mathcal{P}_{\si\rightarrow \si}(t)}{\dr t}\ri| \,\leq  \,2\Delta E  \,\sqrt{\mathcal{P}_{\si\rightarrow \si}(t)\lf(1-\mathcal{P}_{\si\rightarrow \si}(t)\ri)} \, .
\ee
This is formally identical to relation~(\ref{decineq}) for unstable particles. Note that $\si^2_N$ is proportional the \emph{linear entropy} which quantifies the dynamical flavor entanglement of the state \eqref{pontstate}, cf.~\cite{flaventang}. Moreover, (\ref{neutunqm}) is related with
the Wigner--Yanase skew-information~\cite{WigYan}  which reduces to the standard variance on pure states~\cite{Luo}.

If we consider $\mathcal{P}_{\si\rightarrow \si}(t)$ in the interval $0 \leq t \leq t_{1min}$,
where $t_{1min}$ is the time when $\mathcal{P}_{\si\rightarrow \si}(t)$ reaches the first minimum,
this is a monotonically decreasing function \cite{Bil}. In other words, if we try to reveal neutrinos in
processes with time scales much smaller than oscillation time, they can be thought as unstable particles.
In fact, in this time interval, we can regain, by integration, an expression analogous to (\ref{integratedte}).

However, a general inequality, not restricted to a particular
time interval can also be obtained from (\ref{simpdis}). Using the triangular inequality and integrating both
sides from $0$ to $T$, we get
\be
\mbox{\hspace{-2mm}}\Delta E \, T  \ \geq  \ \int^T_{0} \!\! \dr t \, \lf|\frac{\dr\mathcal{P}_{\si\rightarrow \si}(t)}{\dr t} \ri| \, \geq  \,\lf|\int^T_{0} \!\! \dr t \, \frac{\dr\mathcal{P}_{\si\rightarrow \si}(t)}{\dr t} \ri| \, .
\ee
Therefore, one finds
\be \label{etrel}
\Delta E\,  T \geq \mathcal{P}_{\si\rightarrow \rho}(T)  \, ,  \quad \si \neq \rho  \, ,
\ee
with $
\mathcal{P}_{\si\rightarrow \rho}(t) \ = \  1-\mathcal{P}_{\si\rightarrow \si}(t) $.
For $T=T_h$, we finally have
\be
\Delta E\, T_h \geq \ha \, ,
\ee
which is even stronger than (\ref{bhatin}) and, in addition, it has  Heisenberg-like lower bound.

{\em {TEUR for neutrino oscillations in QFT}.}~---~
Let us now consider a full QFT treatment of TEUR.
We have seen that these relations are a consequence of the non conservation
of the number of neutrinos with  definite flavor. However, we used basically a quantum
mechanical treatment, having approximated the flavor neutrino states with the simple
expression~(\ref{ponte1}). One can check that these are not
eigenstates of the flavor charges~(\ref{QflavLept}). True flavor eigenstates can be explicitly constructed as
\be \label{bvflavstate}
|\nu^r_{\G k,\si}\ran \ = \ \al^{r\dag}_{\G k,\si} |0\ran_{e ,\mu}  \, .
\ee
where flavor operators and vacuum are taken at reference time $t=0$.
The corresponding oscillation
formula can be found by taking the expectation value of the flavor charges~\cite{BHV99}
\be
\mathcal{Q}_{\si\rightarrow \rho}(t) \ = \ \lan Q_{\nu_\rho}(t) \ran_\si \, ,
\ee
where $\langle \cdots\rangle_\si = \lan \nu^r_{\G k,\si}| \cdots |\nu^r_{\G k,\si}\ran$, which gives
\bea \label{oscfor}
&& \mbox{\hspace{-4mm}}\mathcal{Q}_{\si\rightarrow \rho}(t)  =   \sin^2 (2 \theta)\Big[|U_\G k|^2\sin^2\lf(\om_{\G k}^{_-}t\ri)+  |V_\G k|^2\sin^2\lf(\om_{\G k}^{_+}t\ri)\Big]  , \nonumber \\[1mm]
&& \mbox{\hspace{-4mm}}\mathcal{Q}_{\si\rightarrow \si}(t)  =  1 \ - \ \mathcal{Q}_{\si\rightarrow \rho}(t) \, , \quad \si \neq \rho \, ,
\eea
where now $\om_{\G k}^{_{\pm}}\equiv (\om_{\G k,2}\pm\om_{\G k,1})/2$ and $|U_\G k|^2=1-|V_\G k|^2$ .
This formula presents oscillations on two different time-scales: $T_-  =  2\pi/\om_{\G k}^{_-}$, which is the main one, observed also in the standard treatment, and $T_+  =  2\pi/\om_{\G k}^{_+}$, due to the interaction with the flavor vacuum condensate~\cite{qftmixing}.

In analogy with the above QM treatment for the number operators,
non-conservation of the flavor-charges leads to a particular form of the QFT-based
TEUR. This is because in this case, lepton charge is a natural candidate for a ``clock observable''. 
By employing the fact that
\be
\lf[Q_{\nu_\si}(t) \ , \, H\ri] \ = \ i \, \frac{\dr Q_{\nu_\si}(t)}{\dr t} \ \neq \ 0 \, ,
\ee
we find the {\em flavor--energy} uncertainty relation
\be \label{neutun}
\sigma_H \, \sigma_Q \ \geq \ \frac{1}{2}\lf|\frac{\dr \mathcal{Q}_{\si\rightarrow \si}(t)}{\dr t}\ri|.
\ee
Moreover, one can verify that
\bea \non
\sigma^2_Q & = & \lan Q^2_{\nu_\si}(t)\ran_\si \ - \ \lan Q_{\nu_\si}(t)\ran_\si^2 \\[1mm] \label{varq}
& = & \ \mathcal{Q}_{\si\rightarrow \si}(t)\lf(1-\mathcal{Q}_{\si\rightarrow \si}(t)\ri) \, .
\eea
Eq.~\eqref{varq} quantifies dynamical (flavor) entanglement for neutrino states in QFT, cf. Refs.~\cite{varent,qftflaventang}). 
This should be compared to results (\ref{decineq}) and (\ref{neutunqm}).  By analogy with (\ref{simpdis}), one has
\be
\lf|\frac{\dr \mathcal{Q}_{\si\rightarrow \si}(t)}{\dr t}\ri| \ \leq \ \Delta E \, .
\ee
From (\ref{neutun}) we arrive at the Mandelstam--Tamm TEUR in the form
\be \label{etq}
\Delta E \,T \ \geq\  \mathcal{Q}_{\si\rightarrow \rho}(T)  \, ,  \quad \si \neq \rho .
\ee
When $m_i/|\G k|\rightarrow 0$, i.e. in the relativistic case, we get
\bea \label{firstapprox1}
|U_\G k|^2  \approx  1 \  -  \ \varepsilon(\G k)  \, , \;\;\;\;\;
|V_\G k|^2  \approx  \varepsilon(\G k)  \, ,
\eea
with $\varepsilon(\G k)   \equiv {(m_1-m_2)^2}/{4 |\G k|^2}$.
In the same limit
\be
\om_{\G k}^{_-} \ \approx \ \frac{\delta m^2}{4 |\G k|}\ = \ \frac{\pi}{L_{osc}} \, , \qquad \om_{\G k}^{_+} \ \approx \ |\G k| \, ,
\ee
where  $\delta m^2\equiv m_2^2-m_1^2$  and $L_{osc}\equiv 4\pi |\G k| / \delta m^2$.
Therefore, at the leading order (ultra-relativistic case) $|U_\G k|^2 \rightarrow 1$, $|V_\G k|^2 \rightarrow 0$. In this limit, 
the standard oscillation formula \eqref{pontosc} is recovered
\bea  \label{stoscfor}
\mathcal{Q}_{\si\rightarrow \rho}(t)& \approx & \sin^2 (2 \theta)\sin^2\lf(\frac{\pi L}{L_{osc}}\ri) \,,  \quad \si \neq \rho,
\eea
where  we put $t \approx L$. The RHS of (\ref{stoscfor}) reaches its maximum at $L=L_{osc}/2$ and the inequality (\ref{etq}) reads
\be \label{condne1}
\De E \ \geq \ \frac{2 \sin^2(2\theta)}{L_{osc}} \, .
\ee
Note that because the Hamiltonian is time-independent, so is $\De E$. In particular, the relation~(\ref{condne1}) applies in the interaction vertex.
Inequalities of the form (\ref{condne1})
are well-known in literature and are usually interpreted as conditions of
neutrino oscillations~\cite{Kayser,NeutPheno, Bil}.

Having based our derivation on exact flavor states and charges, we can see the above relations in a new light:
From the inequality~(\ref{condne1}) we  infer that flavor neutrinos have an inherent energy uncertainty
which represents a  bound for future experiments.
%
%
In order to clarify this statement, note that (\ref{ineqrep}) implies that
\be \label{neutort}
\lim_{V \rightarrow \infty}\lan \nu^r_{\G k,i}| \nu^r_{\G k,\si}\ran \ = \ 0 \, , \qquad  i=1,2 \, ,
\ee
i.e. flavor neutrino state, which is produced in a charged current weak decays,
cannot be written as a linear superposition of single-particle mass eigenstates.
The orthogonality condition~(\ref{neutort}) does not hold for the standard flavor states~(\ref{ponte1}), where
%
$\lim_{V \rightarrow \infty}\lan \nu^r_{\G k,1}| \nu^r_{\G k,e}\ran_P \ = \ \cos \theta$.
%
This contradiction is resolved by observing that
\be
\lim_{m_i/|\G k|\rightarrow 0}\,\,\lim_{V \rightarrow \infty} \ \neq \ \lim_{V \rightarrow \infty}\,\, \lim_{m_i/|\G k|\rightarrow 0} \, ,
\ee
which means that the ultra-relativistic limit cannot be taken once the ``thermodynamical'' QFT limit is performed,
but has to be considered just as QM approximation, which does not hold for systems with an infinite number of
degrees of freedom.  Eq.~(\ref{neutort}) should be thus understood as
\begin{eqnarray}
&&\mbox{\hspace{-9mm}}\lan \nu^r_{\G k,i}| \nu^r_{\G k,\si}\ran \! =\! \ {}_{1,2}\lan 0_{\G k}|\al^r_{\G k,1} \al^{r \dag}_{\G k,e}|0_{\G k}\ran_{e,\mu}   \prod_{\G p \neq \G k}\!\! {}_{1,2}\lan 0_{\G p}|0_{\G p}\ran_{e,\mu}  .
\end{eqnarray}
%

Let us now consider corrections beyond the ultra-relativistic limit.
The exact oscillation formula (\ref{oscfor}) reduces in the next-to-leading relativistic order to~\cite{Lee}
%
%
\begin{eqnarray}
\mathcal{Q}_{\si\rightarrow \rho}(t) &\approx& \sin^2 (2 \theta) \Big[\sin^2\lf(\frac{\pi t}{L_{osc}}\ri) \lf(1 - \varepsilon(\G k) \ri)  \nonumber \\
&& +   \ \varepsilon(\G k)  \sin^2\lf(|\G k|t\ri)\Big] \, , \quad \si \neq \rho \, .
\end{eqnarray}
By setting $~T= L_{osc}/2$, the relation~(\ref{etq}), can be  written as
\be
\De E  \ \geq \ \frac{2 \, \sin^2 2 \theta}{L_{osc}} \, \lf[1  -  \varepsilon(\G k) \,  \cos^2\lf(\frac{|\G k|L_{osc}}{2}\ri)\ri] \, ,
\ee
i.e. the bound on the energy is lowered with respect to \eqref{condne1}.
For neutrino masses~\footnote{This values for neutrino masses are taken from~\cite{pdg},
in the case of inverted hierarchy.}: $m_1=0.0497 \, {\rm eV}$, $m_2=0.0504 \, {\rm eV}$,
and $|\G k|= 1 \, {\rm MeV}$, then $\varepsilon(\G k) = 2 \times 10^{-19}$.

On the other hand, in the non-relativistic regime where the pure QFT effects (such as interactions with the vacuum)
are relevant, the full oscillation formula simplifies. To this end we consider, e.g. $|\G k|= \sqrt{m_1 m_2}$. In this case,
\bea
|U_\G k|^2 & = & \ha \ +\ \frac{\xi}{2} \ = \  1-|V_\G k|^2 \, , \\[2mm]
\xi & = & \frac{2\sqrt{m_1 m_2}}{m_1+m_2} \, ,
\eea
and we can rewrite~(\ref{etq}) as
\bea
\De E \, T & \geq & \frac{\sin^2 2 \theta}{2} \, \Big[1 - \,  \cos \lf(\tilde{\om}_{1}T\ri)\cos \lf(\tilde{\om}_{2}T\ri)  \nonumber \\
 && - \, \xi  \sin \lf(\tilde{\om}_{1}T\ri)\sin \lf(\tilde{\om}_{2}T\ri)\Big] \, ,
\eea
with $\tilde{\om}_j = \sqrt{m_j(m_1+m_2)}$. To compare it with the ultra-relativistic case,
we take $T=\tilde{L}_{osc}/4$, with $\tilde{L}_{osc}=4\pi\sqrt{m_1 m_2}/\de m^2$, obtaining
\bea
\De E  & \geq & \frac{2\sin^2 2 \theta}{\tilde{L}_{osc}} \ \lf(1-\chi\ri) \, .
\eea
Here
\begin{eqnarray}
\chi &=& \xi \, \sin \lf(\tilde{\om}_{1}\tilde{L}_{osc}/4\ri)\sin\lf(\tilde{\om}_{2} \tilde{L}_{osc}/4\ri)\nonumber \\
&&+ \ \cos \lf(\tilde{\om}_{1}\tilde{L}_{osc}/4\ri)\cos\lf(\tilde{\om}_{2} \tilde{L}_{osc}/4\ri) \, .
\end{eqnarray}
Substituting the same values as above, for neutrino masses, we obtain $\chi=0.1$, i.e. the original bound on energy decreased by $10\%$.

{\em {Conclusions}.}~---~
By identifying the flavor charges obtained via Noether's theorem and energy as incompatible observables, we have derived  flavor-energy uncertainty relations.  Taking the non-conserved flavor charges as a ``clock observables'', we arrived at the Mandelstam--Tamm version of TEUR, in a full QFT framework.
In the ultra-relativistic regime our results reproduce the standard conditions for neutrino oscillations from Refs.~\cite{Kayser,NeutPheno, Bil}, thus incorporating the achievements of Ref.~\cite{Bil}.

Unlike Ref.~\cite{Bil}, our result is valid for all times and energy scales and improves the bounds of Ref.~\cite{Bhat}.
We have interpreted TEUR for flavor neutrinos as representing fundamental bounds on energy-variances. This interpretation is drawn in a close analogy with the case of unstable particles, where a notion of a sharp mass is not natural, and only mass (energy) distributions are measurable.

We would like to stress that the reason for which the results here obtained generalize in a natural way the usual QM results, 
resides in the fact that the QFT flavor neutrino states are defined as eigenstates of the flavor charges. This is a nontrivial step which is possible because of the unitary inequivalence of the Hilbert spaces for neutrinos with definite masses and those with definite flavor~\cite{qftmixing}.

Let us remark that our study naturally correlates with the research line of Refs.~\cite{flaventang,qftflaventang,alok} where neutrino oscillations have been studied from a quantum information perspective.
Finally, we note  that analogous analysis can be applied to boson mixing, such as for $K^0$, $D^0$ or $B^0$ mesons.


\begin{acknowledgements}
{\em {Acknowledgments.}}~---~P.J. acknowledges support from the Czech Science Foundation (GA\v{C}R), Grant 17-33812L.
\end{acknowledgements}

\end{document}